\documentclass[runningheads]{llncs}

\usepackage{amsmath}
\usepackage{amssymb}
\usepackage{graphicx}
\usepackage{bbm} 

\usepackage{url}
\usepackage{color}

\usepackage[colorinlistoftodos,prependcaption,textsize=tiny]{todonotes}
\usepackage[bookmarks, colorlinks, breaklinks]{hyperref}

\makeatletter
\def\blfootnote{\xdef\@thefnmark{}\@footnotetext}
\makeatother

\newcommand{\Var}{\ensuremath{\mathrm Var}}

\begin{document}

\title{They may look and look, yet not see:\\
BMDs cannot be tested adequately}
\titlerunning{They may look and look, yet not see}

\author{
Philip B. Stark$^{1,*}$
\and
Ran Xie$^{2,*}$
}

\institute{
University of California, Berkeley
 \and
Stanford University
}

\maketitle

\blfootnote{\textsuperscript{*}Authors listed alphabetically.}

\begin{abstract}
    Bugs, misconfiguration, and malware can cause ballot-marking devices (BMDs) to print incorrect votes.
    Several approaches to testing BMDs have been proposed.
    In \emph{logic and accuracy testing} (LAT) and \emph{parallel} or \emph{live} testing, auditors input known test
    votes into the BMD and check whether the printout matches.
    \emph{Passive} testing monitors the rate at which voters ``spoil'' BMD printout, on the theory that if BMDs malfunction, the rate will increase
    noticeably.
    We provide lower bounds that show
    that these approaches cannot reliably detect outcome-altering problems, because:
    (i)~The number of possible voter interactions with BMDs is enormous, so testing interactions uniformly at random is hopeless.
    (ii)~To probe the space of interactions intelligently requires an accurate model of voter behavior, but because the space of interactions is so large, building a sufficiently accurate model requires observing an enormous number of voters in every jurisdiction in every election---more voters than there are in most U.S.\ jurisdictions.
    (iii)~Even with a perfect model of voter behavior, the required number of tests exceeds the number of voters in most U.S.\ jurisdictions.
    (iv)~An attacker can target interactions that are intrinsically expensive to test, e.g., because they involve voting slowly; or interactions
    for which tampering is less likely to be noticed, e.g., because the voter uses the audio interface.
    (v)~Whether BMDs misbehave or not, the distribution of spoiled ballots is unknown and varies by election and possibly by ballot style: historical data do not help much. Hence, there is no way to calibrate a threshold for passive testing, e.g., to guarantee at least a 95\% chance of noticing that 5\% of the votes were altered, with at most a 5\% false alarm rate.
    (vi)~Even if the distribution of spoiled ballots were known to be Poisson, the vast majority of jurisdictions do not have enough voters for passive testing to have a large chance of detecting problems but only a small chance of false alarms.
\end{abstract}


\section{Introduction}\label{introduction-why-test-bmds}

BMDs print votes, often as barcodes or QR codes, together with a human-readable 
text summary (some BMD printout resembles a hand-marked paper ballot, HMPB).
Jurisdictions including the U.S.\ state of Georgia, Los Angeles County, California, and Philadelphia, Pennsylvania, recently purchased 
BMDs for all in-person voters to use.

Bugs, misconfiguration, or malware can make
the printed votes and QR codes differ from each other and from the voter's selections.
Some have argued that this does not compromise election integrity
because voters have the opportunity to inspect 
BMD printout and to start over 
if the printout does not match their intended selections;
and that since voters can make mistakes hand-marking ballots,
HMPBs are no more secure or reliable than BMD printout \cite{quesenbery18}.
We find those arguments unpersuasive:
\begin{itemize}
    \item In some jurisdictions, the official
record of the vote for counts and recounts is the QR code, which voters cannot check.\footnote{%
   See \url{https://rules.sos.ga.gov/gac/183-1-15-.03?urlRedirected=yes&data=admin&lookingfor=183-1-15-.03} (last visited
   5 May 2022).
   Audits in Georgia rely on the human-readable text, but legally cannot correct
   outcomes, and are conducted only for one contest every two years.
}
   \item The arguments equate holding voters responsible for their own errors with holding 
voters responsible for the overall security of the system 
\cite{
appelEtal20,appelStark20}.
    \item Most voters \emph{do not} inspect BMD printout
\cite{demilloEtal18,bernhardEtal20,haynesHood21}.
Those who do rarely detect actual errors \cite{bernhardEtal20,kortumEtal21}.
To reliably detect errors entails voters taking 3--6 \emph{minutes} to compare a written slate of 
candidates with the printed selections \cite{kortumEtal22},
but voters generally spend less than 3
\emph{seconds} reviewing BMD printout \cite{demilloEtal18,haynesHood21}.
    \item If a BMD misprints a voter's selections, 
only the voter can get evidence of the problem: elections conducted using
BMDs are not \emph{contestible} \cite{appelEtal20}.
    \item If BMDs misbehave, there is no way to determine the correct election outcome because there is no 
trustworthy paper record of the vote:
BMDs are not \emph{strongly software independent} \cite{rivest08}.
\end{itemize}
Concerns about BMDs are not merely hypothetical: BMDs 
have caused 
scanners to fail to count votes accurately, to allow voters to vote, and to present 
all voting options to voters,
even after passing LAT
\cite{eac22,shortellTatu19,SneedLA20,ZetterLA20,PrevitiNH20,harte20,cillizza20,MehrotraNH19,fowler20,riggall22}.

BMD advocates also claim BMDs eliminate ambiguous marks, prevent overvotes, 
and warn about undervotes (e.g., \cite{quesenbery18}).
But that presumes BMDs function correctly; the rate of truly ambiguous handmade marks is minuscule \cite{appelEtal20};
and
precinct-based optical scanners also protect against undervotes and overvotes (the Voluntary Voting Systems Guidelines, VVSG, require it).\footnote{%
Such protection has been required since VVSG 1.0;
see section 2.3.3.2 of \cite{vvsg10}.
}
Regardless,
elections conducted using BMDs are not trustworthy unless there is a way to 
ensure that BMD misbehavior did not change any outcome.
(If the paper trail itself is not trustworthy, risk-limiting audit procedures do not help because even an accurate full hand count 
may not reveal who really won.)
Elections---and hence BMDs---need to be protected against malicious, technically capable attackers,
such as nation states.\footnote{%
The U.S.\ Senate Intelligence Committee, the Department of Homeland Security, and the FBI concluded that Russian
state hackers attacked U.S.\ elections in 2016 \cite{USSenateIntel18}.
}
If testing has a high chance of detecting that an outcome was altered by a skilled attacker, it also protects against 
misconfiguration and bugs---which attackers could mimic.

\section{Prior work}
Vulnerabilities of particular BMDs are discussed in depth in expert declarations by J.\ Alex Halderman in \emph{Curling et al. v. Raffensperger et al.}.
Theoretical vulnerabilities of various BMD designs are discussed in
\cite{appelEtal20}.
\cite{wallach20,gilbert19} discuss testing BMDs; here, we quantitatively investigate their heuristic claims.
Three approaches to testing BMDs have been proposed: pre-election logic and accuracy testing (LAT), 
``live'' or ``parallel'' 
testing during the election, and ``passive'' testing
by monitoring the spoiled ballot rate. 
In LAT and parallel testing, auditors make selections on a BMD then
check whether the printout accurately reflects those selections.
The primary difference is that LAT happens before the election and parallel testing happens during the election.
\emph{Passive} testing uses the spoiled ballot rate: if more voters than usual request a do-over,
that might be because the machines are malfunctioning.

\section{How much testing is enough?}\label{how-much-testing-is-enough}

If the paper trail accurately reflects who won, accurate full hand counts and risk-limiting audits (RLAs) can catch and correct
wrong outcomes.
Here, we study whether testing can establish with high confidence that a paper trail printed by BMDs
accurately reflects who won.
If not, a recount need not show who really won, and a genuine RLA is impossible. 

\subsection{Threats and defenses}
We make the following assumptions about BMD threats and defenses:
\begin{enumerate}
    \item Attackers seek to alter the outcome of one or more contests
without being detected. 
    (Some might want to be detected, to undermine public confidence.)
    \item Attackers know the testing strategy. This does not preclude the possibility that the strategy will be adaptive or have
a random element.
    \item Attackers have access to the state history of each BMD, including votes,
machine settings, etc.; auditors do not.
    \item Attackers have an accurate model of voter behavior in past elections, including political preferences, voting speed,
BMD settings, and so on; auditors generally do not, because it would require monitoring voters illegally.
    \item Auditors seek to ensure that if any outcome is altered, there is a 
       high chance of detecting it, while keeping the chance of false alarms small.
    \item Auditors do not know which contest(s), if any, were altered.
    \item Auditors must obey the law and protect voter privacy.
\end{enumerate}

\subsection{Jurisdiction sizes, contest sizes, and margins}\label{jurisdiction-sizes-contest-sizes-and-contest-margins}

U.S.\ elections are typically administered by 
counties, townships, or other political units smaller than states.
A \emph{ballot style} corresponds to the collection of contests a given voter is eligible to vote in.
Typically in the U.S., some contests are on only
a fraction of ballot styles in a jurisdiction, in part because
many small political units have elections for various offices and measures.
Many contests of all sizes are decided by small margins.
For instance, in Georgia, U.S., the reported 
margin in the 2020 presidential election was about 0.2\%.

Few votes need to be changed to
alter the outcome of small contests and contests with small margins.
Conversely, the number of voters in a jurisdiction is an upper bound on the number of passive tests that can be performed 
and on the sample size to ``learn'' voter behavior for efficient parallel testing.
Thus, jurisdiction size is an important constraint on BMD testing.
Since ballot layout, contests, equipment, demographics,
political preferences, and other variables vary across and within jurisdictions and malware could affect only some equipment or ballot styles, 
it is not possible to pool data
across jurisdictions to get more power.

Changing votes on 1\% of ballots in a jurisdiction can
alter the margin of a jurisdiction-wide plurality contest by 2\% if
there are no undervotes or invalid votes in that contest.
If the undervote rate
is 30\%, then changing votes on 1\% of the ballots can change the margin
by \(0.02/0.7 = 2.9\%\).
If a contest is only on 10\% of the ballots and the undervote rate in the contest is 30\%, 
altering the votes on 1\% of ballots could change the margin in that contest by nearly
29\%.

As of 2020, only
1,629 U.S.\ cities had populations of 100,000 or more, of over 81,363
incorporated places \cite{census20}. 
The 2020 median
population of U.S.\ incorporated areas is 1,201, so about half
of the 81,363 incorporated places have turnout of 1,201 or fewer voters.
Thus, an attacker does not have to change many votes to alter the outcome of a
typical contest for an elected official in a U.S.\ city or incorporated
township.
According to \cite{EAVS20} the 2020
median turnout in the 6,405 U.S.\ counties with recorded active voter data was 4,470 voters, and turnout was less
than 11,500 voters for more than 2/3 of jurisdictions. 
In 65.5\% of
states, more than 50\% of counties have fewer than 30,000 active voters.
In 85.5\% of states, more than 50\% of counties have fewer than 100,000
active voters.

\begin{figure}
\begin{minipage}{0.35\textwidth}
        \centering
\centering
\includegraphics[width=1.9in]{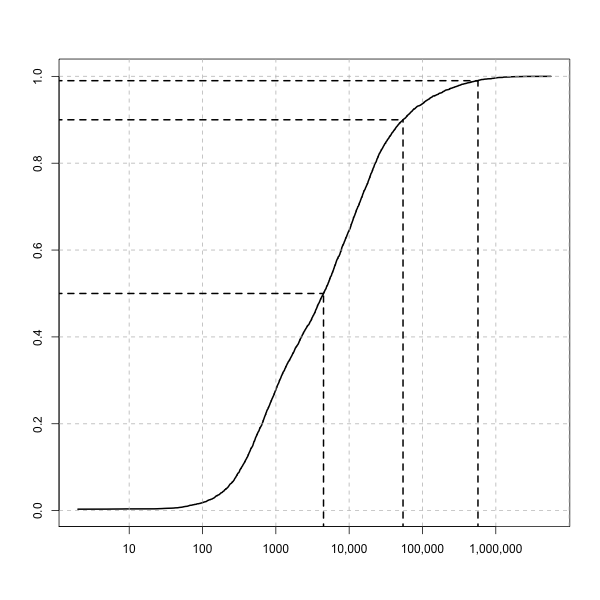}
\caption{2020 turnout by jurisdiction in 3073
counties \cite{EAVS20}. 
Turnout was below 10,000 in $\approx$50\% of counties.}\label{fig:cdf}
\end{minipage}
\hfill
\begin{minipage}{0.55\textwidth}
\centering
\includegraphics[width=3.5in]{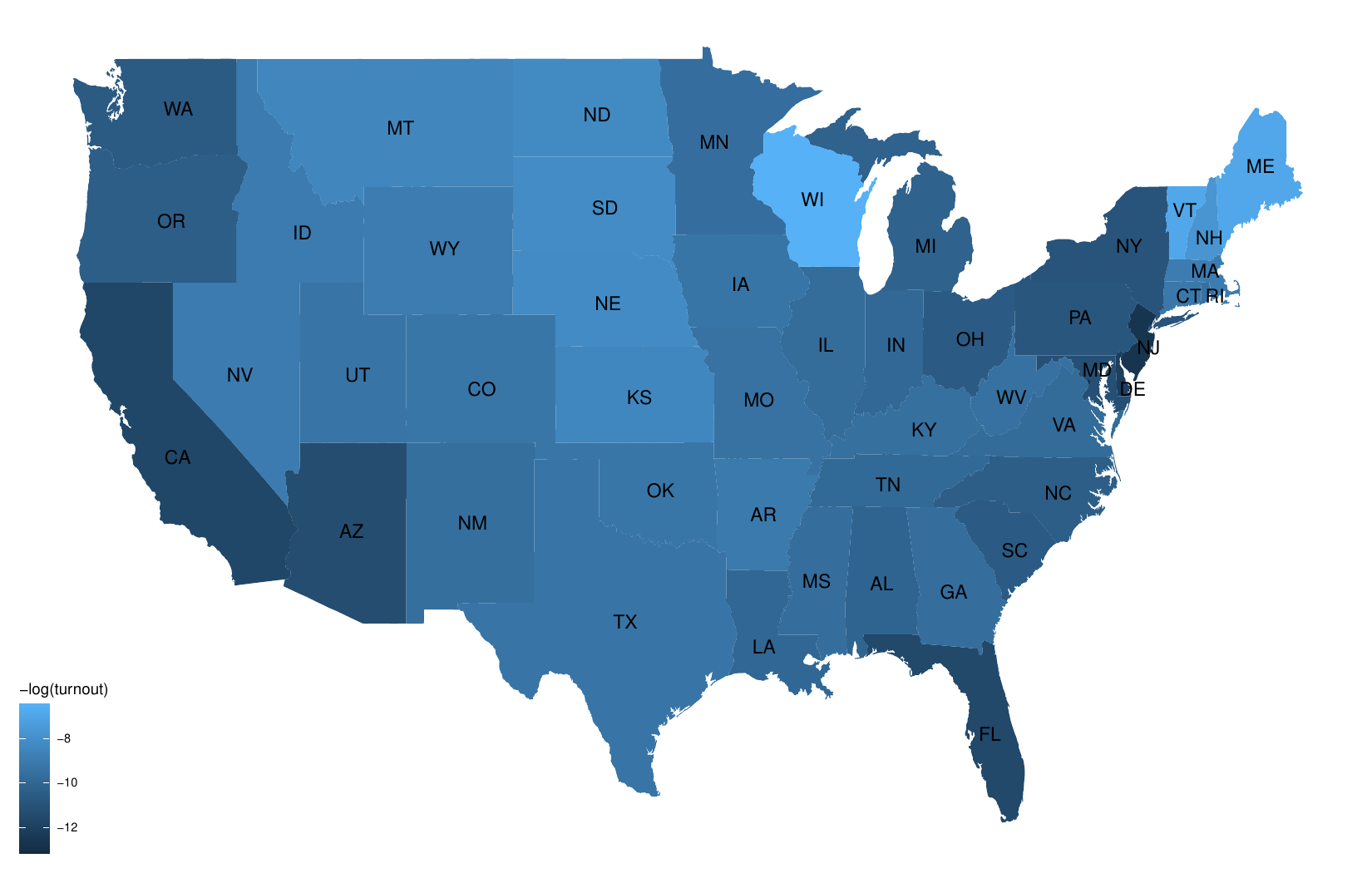}
\caption{Median 2020 turnout by jurisdiction in the U.S. \cite{EAVS20}}\label{fig:heatmap}
\end{minipage}
\end{figure}

\subsection{Voting transactions}\label{voting-transactions}

We shall call a voter's interaction with a BMD a \emph{voting
transaction} or \emph{transaction}. 
Transactions are characterized by many variables, including:
\begin{itemize}
\item
  when the transaction starts
\item
  time since the previous voter finished (a measure of polling-place congestion)
\item
  number of transactions before the current transaction
\item
  the voter's sequence of selections and revisions of selections
\item
  the time to make each selection before
  taking another action
\item
  whether the voter looks at every page of options in each contest
\item
  the time the voter spends reviewing and revising selections
\item
  precisely where voter touches the screen
\item
  BMD settings, including font size, language, use of audio, volume, tempo, pausing, rewinding, use of the sip-and-puff interface,
  inactivity warnings
\end{itemize}
Table~\ref{tbl:tab-dimension} lists some of the variables and the number of values they can take.\footnote{%
  Table~\ref{tbl:tab-dimension} assumes all contests are ``vote-for-one.'' 
  Ranked-choice voting and multi-winner plurality contests yield more possible transactions. 
  Continuous variables were binned into a few options. 
VVSG 1.1  \cite{VVSG15}  requires: 
(a)~Alternative language
access is mandated under the Voting Rights Act of 1975, subject to
certain thresholds (e.g., if the language group exceeds 5\% of the
voting age population). 
(b)~The voting system shall provide the voter
the opportunity to correct the ballot for either an undervote or
overvote before the ballot is cast and counted. 
(c)~An Acc-VS with a
color electronic image display shall allow the voter to adjust the
color saturation throughout the transaction while preserving
the current votes. (d) At a minimum, two alternative display options
listed shall be available: 1) black text on white background, 2) white
text on black background, 3) yellow text on a black background, or 4)
light cyan text on a black background. 
(e)~A voting system that uses
an electronic image display shall be capable of showing all
information in at least two font sizes. (f)~The audio system shall
allow the voter to control the volume throughout the voting
transaction while preserving the current votes. 
(g)~The volume shall
be adjustable from a minimum of 20dB SPL up to a maximum of 100 dB
SPL, in increments no greater than 10 dB. 
(h)~The audio system shall
allow the voter to control the rate of speech throughout the voting
transaction while preserving the current votes. 
(i)~The range of
speeds supported shall include 75\% to 200\% of the nominal rate.
} 
The huge number of possible transactions helps
an attacker pick a subset large enough to change an outcome
but that auditors are unlikely to probe. 
\begin{table}
\centering
\tiny
\begin{tabular}{lcc}
Parameter & optimistic & more realistic\\
\hline
Contests & 3 & 20\\
Candidates per Contest & 2 & 4\\
Languages & 2 & 13\\
Time of day & 10 & 20\\
Number of previous voters & 5 & 140\\
Undervotes & \(2^{3}\) & \(2^{20}\)\\
Changed selections & \(2^{3}\) & \(2^{20}\)\\
Review & 2 & 2\\
Time per selection & 2 & \(5^{20}\)\\
Contrast/saturation & - & 4\\
Font Size & 2 & 4\\
Audio Use & 2 & 2\\
Audio tempo & - & 4\\
Volume & 5 & 10\\
Audio pause & - & \(2^{20}\)\\
Audio + video & - & 2\\
Inactivity warning & 2 & \(2^{20}\)\\
\hline
Total combinations & \(6.14 \times 10^6\) &
\(3.4 \times 10^{48}\)
\end{tabular}
\caption{\protect \label{tbl:tab-dimension}Some parameters of BMD transactions
and their number of possible values.}
\end{table}

\section{Passive testing}\label{sec:passive}

Passive testing sounds an alarm if the number of spoiled ballots exceeds some threshold, $t$.
To ensure that passive testing has a false
negative rate (failing to detect altered outcomes)
of at most $X$\%, we need to know that the chance that the number of 
spoiled ballots is greater than or equal to $t$ is at least $X$\% if BMDs altered any outcome.
Conversely, to limit the false
alarm rate to at most $Y$\%, we need to know that the chance 
that the number of spoiled ballots is greater than or equal to $t$ is at most $Y$\% if BMDs function correctly.

Finding such a value of $t$ is impossible in practice because 
the distribution of spoiled ballots may
depend on ballot design, voting rules, the number of contests, and other things that
vary from election to election and place to place---and when BMDs misbehave, also on 
the number of 
altered transactions, and the voters and contests affected.
Hence, to lower-bound the difficulty, we will assume (optimistically) that the number of
spoiled ballots has a Poisson distribution whether BMDs behave correctly or not; but with a rate that depends on the rate 
of altered transactions.
We assume either that 7\% of voters will notice errors and
spoil their ballots, consistent with the findings of \cite{bernhardEtal20}, 
or that 25\% of voters will. 
We consider contest margins of 1\%--5\% and rates of
false positives (false alarms) and false
negatives (failing to notice altered outcomes) of 5\% and 1\%. 
Results are in table~\ref{tbl:tab-passive-05};
software to calculate these numbers is in
\url{https://github.com/pbstark/Parallel19}. 
\begin{table}
\centering
\tiny
\begin{tabular}{cc|rrr|rrr}
 & voter & \multicolumn{3}{|c}{5\% error rate} & \multicolumn{3}{|c}{1\% error rate} \\
 \cline{3-5} \cline{6-8}
 & detection & \multicolumn{3}{|c}{base spoilage rate} & \multicolumn{3}{|c}{base spoilage rate} \\
margin & rate & 0.5\% & 1\%  & 1.5\%  & 0.5\%  & 1\%  & 1.5\% \\
\hline
1\% & 7\% & 451,411 & 893,176 & 1,334,897  & 908,590 & 1,792,330 & 2,675,912\\
& 25\% & 37,334 & 71,911 & 106,627  & 76,077 & 145,501 & 214,845\\
2\% & 7\% & 115,150 & 225,706 & 336,160  & 233,261 & 454,295 & 675,242\\
& 25\% & 9,919 & 18,667 & 27,325  & 20,624 & 38,039 & 55,442\\
3\% & 7\% & 52,310 & 101,382 & 150,471  & 106,411 & 204,651 & 302,864\\
& 25\% & 4,651 & 8,588 & 12,445  & 9,870 & 17,674 & 25,359\\
4\% & 7\% & 30,000 & 57,575 & 85,227  & 61,385 & 116,631 & 171,908\\
& 25\% & 2,788 & 4,960 & 7,144  & 5,971 & 10,312 & 14,681\\
5\% & 7\% & 19,573 & 37,245 & 54,932  & 40,156 & 75,671 & 110,989\\
& 25\% & 1,838 & 3,274 & 4,689& 4,036 & 6,849 & 9,650
\end{tabular}
\caption{\label{tbl:tab-passive-05}
Minimum turnout for passive testing with a 5\% false negative rate to have at
most a 5\% false positive rate 
(cols 3--5) or for
passive testing with a 1\% false negative rate to have at most a 1\% false positive rate (cols 6--8), as a function of the the contest margin
(col~1), the percentage of voters who would 
notice errors (col~2), and the base rate at which voters spoil BMD printout. 
The number
of spoiled ballots is assumed to have a Poisson distribution, with known rate, absent malfunctions. 
Malfunctions increase the rate
by half the margin times the detection rate.}
\end{table}
Combining
Table~\ref{tbl:tab-passive-05}
and
Figure~\ref{fig:cdf} shows that even 
if the
probability distribution of spoiled ballots were known to be Poisson and
the spoilage rate when equipment functions correctly were known
perfectly, in 2020, in 58.2\% of U.S.\ states fewer
than half the counties had enough voters for passive testing to work,
even in county-wide contests, on the assumption that 7\% of voters whose
votes are altered will spoil their ballots.

If turnout is roughly 50\%, jurisdiction-wide contests in jurisdictions
with fewer than 60,000 voters---22 of California's 58
counties in 2020 \cite{EAVS20}---cannot in principle
limit the chances of false positives and false negatives to 5\% for
margins below 4\%, even under these optimistic assumptions.
For contests that involve only part of a jurisdiction, the situation is worse.

\subsection{Targeting vulnerable voters}\label{sec:unlikely}

The analysis above assumes that all voters are equally likely to detect
errors and spoil their ballots. 
But an attacker can use BMD settings, state
history, and session data to target voters who are
less likely to notice problems.

\paragraph*{Voters with visual impairments.}\label{sec:visual}
Approximately 0.8\% of the U.S.\ population is legally blind;
approximately 2\% age 16 to 64 have a visual impairment \cite{NFB19}. 
Current BMDs do not provide
voters with visual impairments a way to check the printout.
If an attacker only alters votes when the voter uses the audio interface or
large fonts, detection may be very unlikely.

\paragraph*{Voters with motor impairments.}\label{sec:motor}
Some BMDs allow voters to print and cast a ballot without looking at it,
for instance the ES\&S ExpressVote® with ``Autocast,'' aka ``permission to cheat'' \cite{appelEtal20}. 
The attacker can change every vote cast using Autocast, with zero
chance of detection.

\paragraph*{Voters who use languages other than
English.}\label{sec:nnes}
U.S.\ law requires some jurisdictions to provide ballots in languages
other than English. 
For instance, Los Angeles County, CA, provides voting materials in 13
languages \cite{lacc20}. 
In 2013,
roughly 26\% of voters in Los Angeles County spoke a language other than
English at home \cite{lacc20}. 
It is our understanding
that BMDs generally print only in English. 
If voters who use a foreign language on the BMD are unlikely to
check the English-language printout, an attacker could change the outcome of
contests with large margins with
little chance of detection.

\paragraph*{Fast and slow voters.}\label{sec:fast-slow}
An attacker can monitor how long it takes voters to make their selections,
whether they change selections, how long they review the summary screen,
etc. 
A voter who spends little time reviewing selections onscreen may
be unlikely to review the printout carefully. 
Conversely, a
voter who takes a very long time to make selections or changes selections
repeatedly might find voting difficult
or confusing and be unlikely to notice errors. 

Thus, it is in the attackers interest to target the
same groups of voters BMDs are supposed to help: voters with visual
impairments, voters with limited dexterity, voters who use a language other than English, 
and voters with cognitive disabilities. 

\subsection{FUD attacks on passive
testing}\label{fud-attacks-on-passive-testing}
Even under ideal circumstances, passive testing does not produce direct evidence of problems; it
does not identify which ballots or contests have errors;
and it does not provide any evidence about whether problems 
changed outcomes. 
Relying on spoiled ballots as a sign of
fraud opens the door to a simple, legal way to undermine elections: encourage
voters to spoil ballots.

\section{LAT and parallel
testing}\label{la-testing-and-parallel-testing}

Suppose that malware alters one or more votes
with probability \(p\), independently across transactions, uniformly across
voters---regardless of the voter's selections or any aspect of 
the transaction that the attacker
can ascertain. 
Then if auditors make \(n\) tests, the chance that the BMD will
alter at least one of the votes in at least one of the tests---and
the attack will be detected---is \(1 - (1-p)^n\).
For \(p=0.01\), \(n=300\) tests would give a 95\% chance of detecting
a problem.

A BMD can handle roughly 140~transactions per day. 
Testing enough to have a 95\% chance of detecting a 1\%
problem on one BMD would leave no time for voters to
use that BMD. 
Even for pre-election LAT, where capacity for
actual voters is not an issue, conducting 300 ``typical'' tests would
take about 25 hours.

If there were a large number of machines known to have been (mis)programmed 
identically, tests
could be spread across them.
But there are many small contests that need to be tested in conjunction with all other
contests that appear on any ballot style that contains them, and there
is no guarantee that all BMDs in a jurisdiction are programmed
identically.

\emph{This threat model is completely unrealistic}. 
An attacker who wants Alice to
beat Bob will not alter votes for Alice: it would needlessly increase the chance 
of detection.
And as discussed in
section~\ref{sec:unlikely}, rather than randomly changing votes for Bob into votes for Alice, an attacker 
can target transactions that auditors are unlikely to probe.

Setting aside specific machines
for testing facilitates a ``Dieselgate'' type attack  \cite{hotten15}, as 
does conducting tests on a schedule, as suggested by
\cite{gilbert19}. 
Tests need to be unpredictable---with respect to the
specific BMDs tested, time, vote
pattern, duration, and other characteristics of voting transactions---or attackers can avoid detection 
by altering only transactions that do not correspond to any test.
There may be pressure to reduce testing when BMDs are busy, to reduce
waiting times. 
Because malware can monitor the pace of
voting, reducing testing when machines are busy makes it
easier to avoid detection.

An attacker need not alter many transactions to change the
outcome of small contests and contests with small margins.
The fewer votes altered, the more tests
required to ensure a large chance of detection.
To test efficiently, tests should sample more common transactions with
higher probability.
Attackers might be able to estimate of the distribution of transactions
using malware installed on BMDs in previous elections, but
testers will not, since it involves tracking voter behavior at a level of detail that
violates voter privacy. See assumptions~4 and 7 and
section~\ref{sec:building-a-model-of-voter-behavior}.

Auditors do not know which contest(s) and candidate(s) are
affected.
To have a large chance of detecting interference, there needs to
be a large chance of testing a transaction the attacker
alters. 
Attackers can target transactions that are intrinsically expensive to test, e.g., 
transactions that take longer than 10 minutes,
transactions in which the voter 
changes some
number of selections, transactions that display the ballot in a language other than English,
transactions that use the audio
interface at a reduced tempo, etc.

\subsection{Lower bounds on the difficulty of parallel testing} \label{sec:oracle-bounds}

We now study an idealized version of parallel testing, where auditors can tell whether a random
sample of BMD printouts accurately show the voters' selections.
Suppose a contest has 4,470 voters, the median jurisdiction turnout in
2020. 
Suppose that malware alters votes in 
23 transactions, which
could change a margin by more than 1\% in a jurisdiction-wide contest.
How many randomly selected printouts would need to be checked to
have at least a 95\% chance of finding at least one with an
error? 
The answer is the smallest \(n\) such that
\begin{equation} \label{eq:without_replacement}
   \frac{4470-23}{4470} \cdot \frac{4469-23}{4469} \cdots \frac{4470-(n-1)-23}{4470-(n-1)} \le 0.05,
\end{equation} i.e., \(n=546\) printouts, about 12.2\% of the transactions,
corresponding to testing each BMD several times per hour.

Conversely, suppose auditors randomly check
13~printouts per day per machine (on average, testing hourly for a 13-hour day, $\approx$9.2\% of BMD capacity). 
To have at least a 95\% chance of detecting that the outcome of a contest with a
1\% margin was altered, there would need to be at least 6,580 voters in
the contest (almost 150\% the median turnout in
jurisdictions across the U.S.), corresponding to 47 BMDs, even under these unrealistically
optimistic assumptions. 

\subsection{Building a model of voter
behavior}\label{sec:building-a-model-of-voter-behavior}

In practice, auditors cannot check whether voters' BMD printout is correct.
Instead of sampling voters' actual transactions in the election, they will have to come up with their own test transactions.
Testing transactions uniformly at random from all possible transactions is doomed because the number of possible 
transactions is so large.
To mimic voters, auditors might consider sampling from $P$, the population distribution of voting transactions, i.e., the fraction of voters who use the BMD in each of the $S = 6.14 \times 10^6$ ways in the optimistic estimate in Table~\ref{tbl:tab-dimension}.
Suppose an attacker wants to change the outcome of a contest with a margin of $m$, expressed as a fraction of ballots cast (rather than as a number of votes).
The attacker only needs to change a fraction \(m/2\) of the transactions to change
the margin by \(m\).
To have probability at least \(1-\alpha\) of
detecting a change to the outcome of any contest with true margin
\(m\), 
auditors must test in a way that has probability at least
\(1-\alpha\) of sampling at least once from every subset of transactions that contains a fraction \(m/2\) of the transactions.
If auditors could sample transactions independently at random from \(P\),
each sample transaction would have probability $1-m/2$ of \emph{not} being one of the altered transactions.
The chance that $t$ randomly selected transactions would not include one that is altered would be $(1-m/2)^t$.
Thus the number of transactions auditors would need to test is the smallest
\(t\) for which 
\begin{equation}
 ( 1-m/2 )^t \le \alpha, \;\mbox{ i.e.,}
\end{equation}
\begin{equation} \label{eq:with_replacement}
t  \ge \frac { \log \alpha }{ \log (1-m/2) }.
\end{equation}
This is essentially equation~\ref{eq:without_replacement} for sampling with replacement; the two are indistinguishable
when $t$ is small compared to the total number of possible transactions.
A key difference is that in equation~\ref{eq:without_replacement}, auditors are sampling from the actual transactions
in the election, while in equation~\ref{eq:with_replacement}, auditors are sampling from a model, the
frequency distribution of of transactions.

In practice, the auditors do not know $P$---they will have to estimate it by monitoring voters.
In reality, this is impossible to do well: (i)~In a given election, $P$ will depend on the particular contests on the ballot and the particular voters who
participate, both of which change from election to election. (ii)~The variables that characterize a voting transaction include the voter's selections and details about how the voter uses the BMD, so collecting the data would violate voter privacy illegally.
To get a sense of the \emph{statistical} difficulty of the problem, we ignore these practical difficulties. 
If auditors could select voters at random (with replacement) and observe in detail how they use the BMD---all the variables in Table~\ref{tbl:tab-dimension}---that would yield independent, identically distributed (IID) draws from \(P\), which could be used to make an estimate, $\hat{P}$.
If $\hat{P}$ differs too much from $P$, no number of tests will suffice, because $\hat{P}$ might estimate that the frequency
of a transaction is zero when in fact it is sufficiently frequent that altering it could change an outcome.
(By assumption~4, above, the attacker knows $P$ and hence can exploit differences between $\hat{P}$ and $P$.)
How many voters would auditors have to observed to ensure (with sufficiently high probability) that $\hat{P}$ is accurate enough 
for parallel testing?

The $L_1$ distance between two distributions bounds the difference in the probability they assign to any set
($|\hat{P}(A)-P(A)| \le \|\hat{P}-P\|_1/2$).
If $\|\hat{P}- P\|_1 \ge m$, there may be a set \(A\) of
transactions for which \(P(A) = m/2\) but \(\hat{P}(A) = 0\), so
changing votes for transactions in \(A\) could alter some margin\footnote{%
The set of undetectable shifts of $m/2$ votes might not include the one that any particular attacker seeks; this bound is
worst-case across hypothetical attackers and distributions of transactions.
}
by \(m\), with zero chance of
detection, no matter how many tests are performed, if the tests are drawn from $\hat{P}$ rather than $P$.

We cannot guarantee that $\|\hat{P} - P\|_1 \le \varepsilon$ with \emph{certainty}, but 
by observing enough randomly selected voters, we can ensure that the chance 
that $\|\hat{P}-P\|_1 > \varepsilon$ 
is at most $\beta$.
If $\alpha \le \beta$, even an infinite number of tests drawn from $\hat{P}$ may not suffice to guarantee chance at least $1-\alpha$
of detecting outcome-changing manipulations.
If $\alpha > \beta$,
to guarantee chance at least $1-\alpha$ of catching an outcome-changing error, the minimum number of tests required
is
\begin{equation}
\min \left \{t :  ( 1+ \varepsilon/2 - m/2 )^t \le \frac{\alpha-\beta}{1-\beta} \right \} .
\end{equation} 

\paragraph*{Minimax lower bounds.}\label{sec:lower_bound_error}
Suppose auditors draw an IID sample of \(n\) transactions from
\(P\), a frequency distribution on \(S\) possible transactions. 
Let \(\mathcal{M}_S\) denote the collection of all
frequency distributions for those transactions.
Then the training sample size \(n\) must be at least large enough to ensure that
the \(L_1\) error of the best estimator $\hat{P}$ is unlikely to exceed
\(\varepsilon\), provided $P \in \mathcal{M}_S$:
\begin{equation} 
\inf_{\hat{P}} \sup_ { P \in \mathcal{M}_S } 
\Pr_P \{ \| \hat{P} - P \|_1 \le \varepsilon \} \ge 1- \beta.
\end{equation} 

\noindent
\emph{Theorem.} (\cite{hanEtal15}) For any
\(\zeta \in \left( 0, 1 \right]\), \begin{eqnarray}
   \inf_{\hat{P}} \sup_{P \in \mathcal{M}_S} \mathbb{E}_P \| \hat{P} - P\|_1
   &\ge & \frac{1}{8} \sqrt{ \frac{eS}{( 1+\zeta) n}}
    \mathbbm{1} \left ( \frac{(1+\zeta)n}{S} > \frac {e}{16}  \right )  \nonumber \\  
    && + \exp \left ( - \frac{2( 1+\zeta)n}{S} \right ) \mathbbm{1} \left ( \frac {( 1+\zeta)n}{S} 
        \le \frac {e}{16}  \right )  \nonumber \\ 
    && - \exp \left ( - \frac{ \zeta^2 n }{ 24 }  \right ) 
       - 12 \exp \left ( -\frac{\zeta^2 S }{ 32( \ln S)^2 }  \right ), \label{eq:thm-minimax}
\end{eqnarray}  where the infimum is over all
\(\mathcal{M}_S\)-measurable estimators \(\hat{P}\).

\noindent
\emph{Lemma.} Let \(X\) be a random variable with variance
\(\Var{X} \le 1\), and let \(\beta \in (0, 1)\). 
If $\Pr \{X \ge \mathbb{E}X + \lambda \} \le \beta$ then \(\lambda \ge -\sqrt{\beta/(1-\beta)}\).

\noindent
\emph{Proof.} Suppose \(\lambda \ge 0\). Then
\(\lambda \ge - \sqrt{\beta/(1-\beta)}\). Suppose \(\lambda < 0\). By
Cantelli's inequality and the premise of the lemma, \begin{equation}
   \beta \ge \Pr \{X \ge \mathbb{E}X + \lambda \} \ge 1 - \frac{\sigma^2}{\sigma^2 + \lambda^2}
   = \frac{\lambda^2}{\sigma^2 + \lambda^2} \ge \frac{\lambda^2}{1 + \lambda^2}.
\end{equation} Solving for \(\lambda\) yields the desired inequality.
\(\Box\).

Now \(0 \le \|\hat{P} - P \|_1 \le 2\), so
\(\Var \|\hat{P} - P \|_1 \le 1\). 
By the lemma, we need  \(\lambda \ge -\sqrt{\beta/(1-\beta)}\) to ensure 
that $  \Pr \{ \|\hat{P}-P \|_1 \ge \mathbb{E}X + \lambda \} \le \beta$.
If \(\| \hat{P}-P \|_1 \ge 2r\),
there can be a set of transactions \(\tau\) such that \(P(\tau) = m/2\)
but \(\hat{P}(\tau) = 0\), so if tests are generated randomly according to
\(\hat{P}\) there is zero probability of testing any transaction in
\(\tau\), no matter how many tests are performed.
Thus if \(\Pr \{ \|\hat{P} - P \|_1 \ge m \} > \alpha\), even an infinite number of tests cannot guarantee
chance at least \(1-\alpha\) detecting that a fraction
\(m/2\) of the transactions were altered, enough to wipe out a margin of \(m\). 
By the
lemma, that is the case if
\(m < \mathbb{E} \| \hat{P} - P \|_1 -\sqrt{\alpha/(1-\alpha)}\), i.e.,
if \(\mathbb{E} \| \hat{P} - P \|_1 > m+\sqrt{\alpha/(1-\alpha)}\).

The theorem gives a family of lower bounds on
\(\mathbb{E} \| \hat{P} - P \|_1\) in terms of \(n\). 
If the lower bound
exceeds \(m+\sqrt{\alpha/(1-\alpha)}\), testing by drawing transactions
from \(\hat{P}\) cannot protect against all outcome-changing errors.
The bound grows with \(S\), the number of possible transactions. 
To be optimistic, we use an unrealistically small value
\(S = 6.14\times 10^6\) (Table~\ref{tbl:tab-lower-bounds}).\footnote{Software
  implementing the calculations is in  \url{https://github.com/pbstark/Parallel19}.
}

To guarantee a 95\% chance of detecting that \(m/2=5\%\)
of transactions were altered, which could change jurisdiction-wide margins by 10\% or more,
the training sample would need to include at
least 1.082~million transactions, even if auditors could conduct an infinite number
of parallel tests. 
That is larger than the turnout in 99.7\% of U.S.\ jurisdictions in 2020 \cite{EAVS20}; it is 
roughly 0.5\% of the U.S.\ voting population.
To guarantee 99\% chance of detecting that 0.5\% of transactions were altered,
which could change jurisdiction-wide margins by
1\% or more, would require observing 3.876~million voters in complete,
privacy-eliminating detail---more than the turnout in 99.9\% of U.S.\ 
jurisdictions in 2020 \cite{EAVS20}, roughly 1.9\% of the U.S.\ voting population.
\begin{table}
\centering
\tiny
\begin{tabular}{cc|rr}
Confidence & Maximum & Altered & Bound \\
Level & Tests & Votes & (millions)\\
\hline
99\% & 2000 & 0.5\% & 3.87\\
&  & 1\% & 3.58\\
 &  & 3\% & 2.69\\
&  & 5\% & 2.09\\
\hline
95\% & 2000 & 0.5\% & 1.67\\
&  & 1\% & 1.59\\
&  & 3\% & 1.31\\
 &  & 5\% & 1.10\\
\hline
99\% & Inf & 0.5\% & 3.73\\
 &  & 1\% & 3.46\\
&  & 3\% & 2.61\\
 &  & 5\% & 2.04\\
\hline
95\% & Inf & 0.5\% & 1.65\\
 &  & 1\% & 1.57\\
 &  & 3\% & 1.29\\
 &  & 5\% & 1.08
\end{tabular}
\caption{\label{tbl:tab-lower-bounds}Lower bound on the sample size (col~4) required to estimate the distribution of voting
transactions well enough to ensure the probability (col~1)
of detecting the manipulation of the fraction of transactions
(col~3) using some number of tests (col~2), if the support of
the distribution of transactions has \(S=6.14\times 10^6\)
points.}
\end{table}

\begin{center}
\includegraphics[width=0.5\textwidth,height=2.5in]{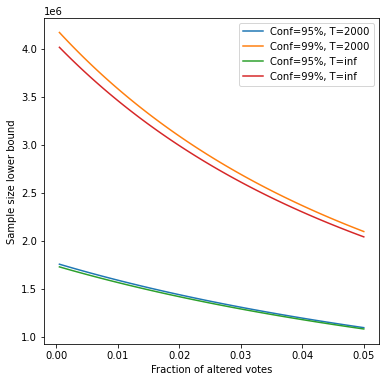}
\end{center}
\begin{figure}[!h]
\caption{Minimum training sample sizes as a function of the fraction of
altered votes.}
\end{figure}

\section{Complications}\label{complications-and-frustrations}
Reality is worse than the optimistic assumptions in our analyses:

\paragraph*{Margins are not known in advance.}
Margins are not known until the election is over, when it is too late to
do more testing if contests have narrower margins than anticipated.
Testing to any pre-defined threshold, e.g., a 95\% chance of
detecting changes to 0.5\% of the votes in any contest, will
not always suffice. 

\paragraph*{Tests have uncertainty.}
If the BMD printout reflects the wrong electoral outcome, a
perfect full manual tally, recount, or risk-limiting audit based on BMD
printout will confirm that wrong outcome.
Suppose one could design
practical parallel tests that had a 95\% chance of sounding an alarm if
BMDs alter $\ge 0.5\%$ of the votes in any contest. 
A reported margin
of 1\% or less in a plurality contest
is below the ``limit of detection'' of such tests.
Would laws require a runoff whenever a reported margin is
below the limit of detection of the tests?



%

\paragraph*{Special risks for some voters.}
As discussed above in Section~\ref{sec:unlikely}, 
BMDs can be used
to selectively disenfranchise voters with disabilities and voters
whose preferred language is not English. 
Indeed, the attacker's best strategy
is to target such voters, in part because
poll workers might more likely to think that complaints by such voters
reflect voter mistakes rather than BMD malfunctions.

\paragraph*{The only remedy is a new election.}
If a BMD is caught misbehaving, it should be removed from service and all BMDs in that jurisdiction
should be investigated. 
But there is no way to determine the correct outcome or which votes were
affected:
BMDs are not \emph{strongly software
independent} \cite{rivest08}.

\section{Conclusion}\label{conclusion}

We show that to protect against outcome-altering BMD malfunctions 
requires orders of magnitude more testing than is feasible.
To our knowledge, no jurisdiction has conducted \emph{any}
parallel testing of BMDs of the kind suggested by \cite{wallach20,gilbert19}, 
much less enough to reliably detect
outcome-changing errors, bugs, or hacks. 

Even if it were possible to test enough to get high confidence that no
more than some threshold percentage the votes were changed in any
contest, fairness would demand a runoff in contests
decided by less than that threshold.

Some BMDs may be the best extant technology for voters with
particular disabilities to mark and cast a paper ballot
independently. 
But many BMDs are poorly designed.
Some have easily exploited security flaws
\cite{appelEtal20} 
and some do not enable voters with common disabilities to
vote independently \cite[pp68--90]{paSoS18}.
To our knowledge, no VVSG-certified BMD system provides a means
for blind voters to check whether the printout matches their intended
selections.

Using BMDs makes elections less trustworthy, less resilient, less transparent, more fragile, and more expensive \cite{appelEtal20,perezMiller19}.
BMDs have failure modes that hand-marked paper ballots do not have,
and lack resilience when failures occur
\cite{appelEtal20}.
BMDs shift the burden of ensuring that voting equipment functions correctly from officials to voters, but do not provide voters any way to prove that they observed problems, if they do; nor can election officials show that outcomes are correct despite any problems that might have occurred
\cite{appelEtal20}.
BMDs undermine the ability of election officials to provide affirmative evidence that outcomes
are correct, the fundamental principle of ``evidence-based elections''
\cite{starkWagner12,appelStark20}.

Voters who use BMDs should be urged to bring a written list of their selections to the polls to check against BMD printout, and to request a fresh ballot if
the printout does not match their intended selections. 
Election officials should track spoiled
BMD printouts.
There should be research on how to encourage voters to check BMD printout and report discrepancies, how to ensure the checks are accurate, and how to ensure that any reported problems are accountably and transparently recorded, addressed, and publicized; these issues also arise in end-to-end cryptographically verifiable (E2E-V) voting systems.
For the foreseeable future, prudent election administration requires
keeping the use of BMDs to a minimum.

\subsubsection*{Acknowledgements}
We are grateful to Yanjun Han and Tsachy Weissman for helpful conversations about minimax $L_1$ estimation,
and to Peter R\o nne and anonymous referees for helpful comments and suggestions.
\bibliography{bib.bib}

\begin{thebibliography}{10}
\providecommand{\url}[1]{\texttt{#1}}
\providecommand{\urlprefix}{URL }
\providecommand{\doi}[1]{https://doi.org/#1}

\bibitem{appelEtal20}
Appel, A., DeMillo, R., Stark, P.: Ballot-marking devices ({BMDs}) cannot
  assure the will of the voters. Election Law Journal  \textbf{19},  432--450
  (2020). \doi{10.1089/elj.2019.0619},
  \url{https://papers.ssrn.com/sol3/papers.cfm?abstract\_id=3375755}

\bibitem{appelStark20}
Appel, A., Stark, P.: Evidence-based elections: Create a meaningful paper
  trail, then audit. Georgetown Law Technology Journal  \textbf{4.2},  523--541
  (2020),
  \url{https://georgetownlawtechreview.org/wp-content/uploads/2020/07/4.2-p523-541-Appel-Stark.pdf}

\bibitem{bernhardEtal20}
Bernhard, M., McDonald, A., Meng, H., Hwa, J., Bajaj, N., Chang, K., Halderman,
  J.: Can voters detect malicious manipulation of ballot marking devices? In:
  41st {IEEE} {S}ymposium on {S}ecurity and {P}rivacy. pp. 679--694. IEEE
  (2020). \doi{10.1109/SP40000.2020.00118}

\bibitem{cillizza20}
Cillizza, C.: How did {G}eorgia get it so wrong (again)? (2020),
  \url{https://www.cnn.com/2020/06/10/politics/georgia-primary-vote-brian-kemp/index.html}

\bibitem{demilloEtal18}
DeMillo, R., Kadel, R., Marks, M.: What voters are asked to verify affects
  ballot verification: A quantitative analysis of voters' memories of their
  ballots. \url{https://ssrn.com/abstract=3292208} (Nov 2018).
  \doi{http://dx.doi.org/10.2139/ssrn.3292208}

\bibitem{fowler20}
Fowler, S.: State outlines fix for error that halted election testing. Georgia
  Public Broadcasting  (2020),
  \url{https://www.gpb.org/news/2020/09/29/state-outlines-fix-for-error-halted-election-testing}

\bibitem{gilbert19}
Gilbert, J.: Ballot marking verification protocol.
  \url{http://www.juangilbert.com/BallotMarkingVerificationProtocol.pdf} (2019)

\bibitem{hanEtal15}
Han, Y., Jiao, J., Weissman, T.: Minimax estimation of discrete distributions.
  In: 2015 IEEE International Symposium on Information Theory (ISIT). pp.
  2291--2295. IEEE (2015)

\bibitem{harte20}
Harte, J.: Exclusive: {P}hiladelphia's new voting machines under scrutiny in
  {T}uesday's elections. Reuters  (2020),
  \url{https://in.reuters.com/article/usa-election-pennsylvania-machines/exclusive-philadelphias-new-voting-machines-under-scrutiny-in-tuesdays-elections-idINKBN2382D2}

\bibitem{haynesHood21}
Haynes, A., III, M.H.: Georgia voter verification study.
  https://s3.documentcloud.org/documents/21017815/gvvs-report-11.pdf (2021),
  last visited 31 October 2021

\bibitem{hotten15}
Hotten, R.: Volkswagen: The scandal explained.
  \url{https://www.bbc.com/news/business-34324772} (2015)

\bibitem{kortumEtal22}
Kortum, P., Byrne, M., Azubike, C., Roty, L.: Can voters detect errors on their
  printed ballots? {A}bsolutely. \url{https://arxiv.org/abs/2204.09780} (2022)

\bibitem{kortumEtal21}
Kortum, P., Byrne, M., Whitmore, J.: Voter verification of ballot marking
  device ballots is a two-part question: Can they? mostly, they can. do they?
  mostly, they don't. Election Law Journal: Rules, Politics, and Policy
  \textbf{September},  243--253 (2021). \doi{10.1089/elj.2020.0632}

\bibitem{lacc20}
{Los Angeles County Clerk}: Multilingual services program (2020),
  \url{https://www.lavote.net/home/voting-elections/voter-education/multilingual-services-program/multilingual-services-program}

\bibitem{MehrotraNH19}
Mehrotra, K., Newkirk, M.: Expensive, glitchy voting machines expose 2020
  hacking risks (2019),
  \url{https://www.bloomberg.com/news/articles/2019-11-08/expensive-glitchy-voting-machines-expose-2020-hacking-risks}

\bibitem{NFB19}
{National Federation of the Blind}: Blind statistics (2019),
  \url{https://www.nfb.org/resources/blindness-statistics}

\bibitem{perezMiller19}
Perez, E., Miller, G.: {Georgia State Election Technology Acquisition: A
  Reality Check}.
  \url{https://trustthevote.org/wp-content/uploads/2019/03/06Mar19-OSETBriefing_GeorgiaSystemsCostAnalysis.pdf}
  (2019)

\bibitem{PrevitiNH20}
Previti, E.: Northampton officials unanimously vote ‘no confidence’ in
  {ExpressVote XL} voting machine (2019),
  \url{https://papost.org/2019/12/20/northampton-officials-unanimously-vote-no-confidence-in-expressvote-xl-voting-machine/}

\bibitem{quesenbery18}
Quesenbery, W.: Why not just use pens to mark a ballot? (2018),
  \url{https://civicdesign.org/why-not-just-use-pens-to-mark-a-ballot/}

\bibitem{riggall22}
Riggall, H.: Up to 157 incorrect ballots cast on first day of early voting,
  {C}obb elections director says. Marietta Daily Journal, Ga  (2022),
  \url{https://news.yahoo.com/157-incorrect-ballots-cast-first-092000088.html}

\bibitem{rivest08}
Rivest, R.: On the notion of `software independence' in voting systems. Phil.
  Trans. R. Soc. A  \textbf{366}(1881),  3759--3767 (October 2008)

\bibitem{paSoS18}
{Secretary of the Commonwealth of Pennsylvania}: Report concerning the
  examination results of election systems and software {EVS} 6012 with {DS200}
  precinct scanner, {DS450} and {DS850} central scanners, {ExpressVote} {HW}
  2.1 marker and tabulator, {ExpressVote} {XL} tabulator and electionware
  {EMS}.
  \url{https://www.dos.pa.gov/VotingElections/Documents/Voting\%20Systems/ESS\%20EVS\%206021/EVS\%206021\%20Secretary\%27s\%20Report\%20Signed\%20-\%20Including\%20Attachments.pdf}
  (2018)

\bibitem{shortellTatu19}
Shortell, T., Tatu, C.: Here's why {N}orthampton {C}ounty's voting machines
  went wrong, county executive says. The Morning Call  \textbf{12 December
  2019} (2019)

\bibitem{SneedLA20}
Sneed, T.: Will {L.A.}'s voting overhaul be an industry disrupter or the next
  election debacle? (2020),
  \url{https://talkingpointsmemo.com/news/will-l-a-s-voting-overhaul-be-an-industry-disrupter-or-the-next-election-debacle}

\bibitem{starkWagner12}
Stark, P., Wagner, D.: Evidence-based elections. IEEE Security and Privacy
  \textbf{10},  33--41 (2012)

\bibitem{census20}
{U.S. Census Bureau}: City and town population totals: 2010-2020 (2020),
  \url{https://www2.census.gov/programs-surveys/popest/datasets/2010-2020/cities/}

\bibitem{vvsg10}
{U.S. Election Assistance Commission}: Voluntary voting systems guidelines 1.0
  (Dec 2005),
  \url{https://www.eac.gov/sites/default/files/eac\_assets/1/28/VVSG.1.0\_Volume\_1.PDF}

\bibitem{VVSG15}
{U.S. Election Assistance Commission}: Voluntary Voting System Guidlines
  version 1.1. U.S. Election Assistance Commission (2015),
  \url{https://www.eac.gov/sites/default/files/eac\_assets/1/28/VVSG.1.1.VOL.1.FINAL1.pdf}

\bibitem{EAVS20}
{U.S. Election Assistance Commission}: The {U.S.} {E}lection {A}ssistance
  {C}ommission's 2020 election administration and voting survey ({EAVS})
  (2020), \url{https://eavsportal.com/}

\bibitem{eac22}
{U.S. Election Assistance Commission}: {Report of Investigation, Dominion
  Voting Systems D-Suite 5.5-B, Williamson County, Tennessee}.
  \url{https://www.eac.gov/sites/default/files/TestingCertification/EAC_Report_of_Investigation_Dominion_DSuite_5.5_B.pdf}
  (2022)

\bibitem{USSenateIntel18}
{U.S. Senate Intelligence Committee}: Russian targeting of election
  infrastructure during the 2016 election: Summary of initial findings and
  recommendations.
  \url{https://www.burr.senate.gov/imo/media/doc/RussRptInstlmt1-\%20ElecSec\%20Findings,Recs2.pdf}
  (2018), last visited 3 June 2020

\bibitem{wallach20}
Wallach, D.: On the security of ballot marking devices. The Ohio State
  Technology Law Journal  \textbf{16.2},  558--586 (2020)

\bibitem{ZetterLA20}
Zetter, K.: {Los Angeles County's} risky voting experiment (2020),
  \url{https://www.politico.com/news/2020/03/03/los-angeles-county-voting-experiment-119157}

\end{thebibliography}

\end{document}